\documentclass[preprint,aps,eqsecnum,nofootinbib]{revtex4}
\usepackage{epsfig}

\newcommand{\be}{\begin{equation}}
\newcommand{\ee}{\end{equation}}
\newcommand{\bea}{\begin{eqnarray}}
\newcommand{\beas}{\begin{eqnarray*}}
\newcommand{\eea}{\end{eqnarray}}
\newcommand{\eeas}{\end{eqnarray*}}
\newcommand{\ba}{\begin{array}}
\newcommand{\ea}{\end{array}}
\def\ls{\mathrel{\lower4pt\vbox{\lineskip=0pt\baselineskip=0pt
           \hbox{$<$}\hbox{$\sim$}}}}
\def\gs{\mathrel{\lower4pt\vbox{\lineskip=0pt\baselineskip=0pt
           \hbox{$>$}\hbox{$\sim$}}}}

\begin{document}



\title{Sneutrino condensate as a candidate for the hot big bang cosmology}
\author{Anupam Mazumdar~$^{1}$, and A. P\'erez-Lorenzana~$^{2}$}

\affiliation{
$^{1}$~McGill University, 3600 University Road, Montr\'eal,
Qu\'ebec, H3A 2T8, Canada\\ 
$^{2}$~Departamento de F\'{\i}sica,Centro de Investigaci\'on y de Estudios 
Avanzados del I.P.N. Apdo. Post. 14-740, 07000, M\'exico, D.F., M\'exico}


\begin{abstract}
If inflationary paradigm is correct, then it must create conditions
for the hot big bang model with all observed matter, baryons and the
seed perturbations for the structure formation. In this paper we
propose a scenario where the inflaton energy density is dumped into
the bulk in a brane world setup, and all the required physical
conditions are created by the right handed neutrino sector within
supersymmetry. The scalar component of the right handed Majorana
neutrino is responsible for generating the scale invariant
fluctuations in the cosmic microwave background radiation, reheating
the Universe at a temperature~$T_{rh}\leq 10^{9}$~GeV, and finally
generating the lepton/baryon asymmetry, $n_{B}/s\sim 10^{-10}$, with
no lepton/baryon isocurvature fluctuations.

\end{abstract}

\maketitle
\section{Introduction and summary}

Inflation is the only dynamical mechanism which stretches the quantum
fluctuations causally outside the horizon, besides making the Universe
flat, homogeneous and isotropic~\cite{Guth}.  The primordial
perturbations, which enters the Hubble horizon after the end of
inflation, act as a seed for the large scale structures and the
fluctuations in the cosmic microwave background
radiation~\cite{WMAP}. The vacuum dominated inflation leaves the
Universe devoid of any entropy, therefore it is important that any
inflationary paradigm must pave a way to a hot big bang cosmology with
a matching Standard Model (SM) relativistic degrees of freedom.

It is usually believed that the entropy is generated from the decay of
the inflaton energy density, however it is unfortunate that the
inflaton sector remains enigmatic. It is often regarded as a SM gauge
singlet, and in many models an absolute gauge singlet, whose potential
is sufficiently flat to provide enough e-foldings of inflation with
almost scale invariant density perturbations 
In this regard the inflaton potential can be considered as a dark
energy~\footnote{ Often in the literature the dark energy is referred
to the present acceleration, for which we do not have a good
understanding, see~\cite{Carroll}. In this paper we regard the
primordial inflation as a dark energy sector, mainly because we do not
know the inflaton sector at all.} with an yet unanswered puzzle, ``how
does this unknown sector couples to the SM degrees of freedom?''

Indeed it is possible that the inflaton does not couple to the SM
degrees of freedom at all, as suggested in
Refs.~\cite{Enqvist1,Enqvist2,Enqvist3,Lorenzana,Lorenzana1,Brandenberger1}.
In these examples the gauge invariant flat directions of minimal
supersymmetric SM (MSSM), for a review see~\cite{MSSM}, was
responsible for reheating the Universe with the MSSM degrees of
freedom.  In Ref.~\cite{Lorenzana}, we explored similar possibility
with a sneutrino condensate, see also Ref.~\cite{McDonald}.

Although conventionally one relates the spectrum of the density
perturbations to the properties of the inflaton potential, recently,
it has been realized that this is not a necessary condition for a
successful inflationary scenario. In curvaton models the dark energy
induces quantum fluctuations in a field whose energy density during
inflation is negligible, but which may later become dominant
\cite{Sloth,Lyth,Moroi,Others}.

On the other hand, the recent developments in the neutrino experiments
suggest that the neutrinos have non-vanishing masses.  Solar neutrino
deficit is better understood if the electron neutrinos oscillate into
the muon neutrinos controlled by the squared mass difference, $\Delta
m^2_{solar}\sim 7\times 10^{-5}~{\rm eV}^2$, with a large mixing angle
$\tan^2\theta_{solar}\sim 0.5$~\cite{Fukudas}, whereas the atmospheric
neutrino experiments indicate $\nu_{\mu}-\nu_{\tau}$ oscillations with
$\Delta m^2_{atm}\sim 2.5\times 10^{-3}~{\rm eV}^2$ and
$\sin^2(2\theta)\simeq 1$~\cite{Fukudaa}.

If the neutrinos were Dirac particles, unnaturally small Yukawa
couplings of order one part in $10^{11}$ were required in order to
obtain $m_\nu\sim 1$~eV.  The most natural explanation arises if the
neutrinos are Majorana particles along with the see saw mechanism that
involves large right handed neutrino masses~\cite{Seesaw}. One
advantage of this mechanism is that the right handed neutrino mass
breaks $L$ (or $B-L$) quantum number, which can create a lepton
asymmetry. The conversion of the leptonic asymmetry into the baryonic
asymmetry, via active SM sphalerons within a range of $10^{12}~{\rm
GeV}\geq T\geq 100$~GeV, addresses the observed baryon
asymmetry~\cite{Fukugita}.

In a supersymmetric theory, which we will assume in our analysis, the
sneutrino condensate decay can also induce leptogenesis. This scenario
in principle can be tested by the cosmic microwave background
radiation (CMB) via the baryon-isocurvature
fluctuations~\cite{Mazumdar1}.  In our case the sneutrino will decay
into the sleptons and the Higgsinos, thus ensuring reheating into the
MSSM and the SM degrees of freedom.

In Refs.~\cite{Enqvist3,Lorenzana1,Lorenzana}, it was suggested that
the inflaton energy density can be deposited away from the observable
world within a brane world setup. In brane world models the Universe
is assumed to be a four dimensional surface (brane) embedded in a bulk
of $4+n$ dimensions. The $n$-extra dimensions can be either open or
compact (for a review see \cite{Rubakov1}). The standard setup assumes
that the higher dimensional graviton propagates in the bulk, while the
SM gauge fields are stuck on the brane. The bulk geometry could be
warped with an anti-de-Sitter (adS) metric \cite{RS1,RS2}. In these
scenarios the higher dimensional Planck scale $M_{\ast}$ can be
different from the four dimensional Planck mass $M_{p}=2.4\times
10^{18}$~GeV, which are related to each other by the volume of the
extra space. When the bulk is adS, it is possible to localize the zero
mode of the bulk gravitons as a four dimensional graviton from the
brane point of view.

In the background of an adS geometry the quasi-localized zero mode of
a massive bulk field, at a brane position, has a finite life time to
leave the brane~\cite{Rubakov2}. It is equally possible that a brane
field has an exclusive coupling to the bulk degrees of freedom.  In
the latter case also the brane field will eventually decay into the
bulk modes.  Suppose that the inflaton is such a candidate which
couples only to the bulk modes, this could happen due to some
underlying symmetry which prevents it from having renormalizable and
non-renormalizable couplings to the MSSM matter fields, in which case
the question arises; how would the Universe evolve after inflation?

Inflation excites any light scalar fields. Therefore any light degrees
of freedom, lighter than the Hubble expansion rate, will obtain
quantum fluctuations. Such light fields act as a homogeneous
condensate during inflation, see~\cite{MSSM}. The dynamics of the
sneutrino fields can be almost frozen if their masses are lighter than
the Hubble expansion rate. Obviously the dynamics of the lighter ones
will prevail over the others. The fluctuation in the lightest field is
of isocurvature in nature during inflation, however after inflation
the lightest sneutrino condensate dominates the energy density. During
its decay the sneutrino not only generates lepton asymmetry and the
MSSM relativistic degrees of freedom, but also converts isocurvature
perturbations into the adiabatic perturbations. It is therefore the
sneutrino condensate which leads to the hot big bang cosmology and not
the inflaton, the inflaton sector remains the darkest side of the
Universe.

We begin our discussion with an inflaton decaying into the bulk modes.
We then describe the origin of light and heavy Majorana neutrino
masses. We discuss the dynamics of the sneutrino condensate and
explain various implications.


\section{Reheating the bulk modes}

For an example we will work in a 5D brane world setup, where the
bulk is infinitely large with a warped metric
\begin{equation}
\label{metric}
ds^2=g(z)_{MN}dx^M dx^N = \omega^2(z)g_{\mu\nu}dx^{\mu}dx^{\nu}-dz^2\,,
\end{equation}
where $g_{\mu\nu}$ is the four dimensional brane metric, with all the
SM gauge degrees of freedom fixed to the brane located at
$z=0$~\cite{RS1}.  The warp factor in the metric has a form
$\omega(z)=e^{-\kappa|z|}$, where $\kappa$ is a constant that relates the
true gravitational coupling constant of the five dimensional theory,
$G_{5}$, to the low energy effective four dimensional Newton's
constant via $G_N=\kappa~G_{5}$.  In this theory the Planck scale,
$M_{P}=(8\pi G_N)^{-1/2}=2.4\times 10^{18}$~GeV, is derived from the
fundamental gravitational scale, $M_{\ast}=(8\pi G_{5})^{-1/3}$,
through the relationship
 \begin{equation}
 M_\ast^3=\kappa~M_P^2\,.
 \end{equation}
Here onwards we will take $\kappa$, and therefore $M_\ast$, to be
close to the Planck scale~\footnote{For instance, $M_\ast\sim 0.5
M_{p}$. This gives $\kappa\sim 0.1 M_{p}$.}.  In this setup the brane
has a tension, $\sigma = 6\kappa M_\ast^3$, whereas the bulk is an adS
slice with a negative cosmological constant $\Lambda
=-\sigma^2/6M_\ast^3$~\cite{RS2}.

For the purpose of illustration, let us assume that the inflaton is a
true 4D brane field which is homogeneously distributed and dominating
the energy density.  The modified Friedmann equation has a quadratic
dependence on the brane density, $\rho$,
see~\cite{langlois,carlos,Visser,March}. From the point of view of an
effective field theory on the brane, the Hubble expansion rate,
$H=\dot a /a$, where $a$ is the scale factor of the four dimensional
Universe, is determined by
\be H^2={1\over 3 M_P^2}\rho \left(1 +
{\rho\over 2\sigma} \right)
+\frac{\sigma}{2~M_{p}^2}\left(\frac{a_{h}}{a}\right)^4\,.
\label{friedman} 
\ee 
The last term is an interesting one, which needs some explanation.
The adS has an horizon, in order to see that; let us assume that a
point particle is traveling in the bulk, it has a momentum along the
fifth component. Although the particle is moving towards infinity, it
actually takes a finite proper time, $\tau=(\pi/2\kappa)$, to reach
the end of the space, see~\cite{Muck}, hence $z=\infty$ actually acts
as a particle horizon. The boundary conditions at the horizon are
imposed such that nothing comes in from behind the horizon.

The four dimensional Poincar\'e invariance ensures that the coordinate
four momentum, $p_{\mu}$, coincides with the physical momentum on the
brane, but from the point of view of an observer, away from the brane,
$z\neq 0$, the four dimensional momentum gets blue shifted, this means
that the modes which are softer on the brane become harder away from
the brane. This leads to a conjecture that a black hole might form at
the horizon of the adS space~\cite{Visser,March}. In the above
expression, $a_{h}$ is understood as a black hole
horizon~\cite{Visser}. It is always compared with the scale factor
$a(t)$ on the brane.  The second term is a correction due to the brane
position breaking the 5D Lorentz invariance, such that the brane
tension is acting as a source term in 5D.

The main point is to note that the standard Hubble relationship,
$H = \sqrt{\rho/3 M_P}$, follows only for small densities compared to
the brane tension, $\rho\ll \sigma$~\cite{langlois,carlos}, and the
contribution from the last term is sub dominant as long as the scale
factor on the brane position, $a\gg a_{h}$, is greater than the black
hole horizon \cite{March}. With these assumptions thermal history of
the Universe follows similar to the standard cosmology~\cite{Maz}. 
We will always be working at energy scales below the brane tension,
therefore with the standard Friedmann equation.

Let us suppose that the inflaton and the bulk fields carry some global
quantum number, while the brane degrees of freedom do not. In such a
case the inflaton energy will be radiated away into the bulk after the
end of inflation in the form of KK modes, from the four dimensional
point of view. These bulk modes are carrying the momentum along the
fifth dimension, therefore they will simply fly away towards infinity.
For the propose of illustration, let us now assume that the brane
inflaton field, $\phi$, couples to some bulk scalar field through the
coupling~\footnote{The inflaton coupling to the fermions were already
studied in Ref.~\cite{Lorenzana}, here we describe the bosonic interaction.}
\be 
\sqrt{g(z)}\xi~\phi(x) \varphi(x,z)\varphi(x,z)\delta(z)\,,
 \label{bbc}
\ee 
where $\xi$ is a dimensionless coupling constant.  This coupling
induces a large decay rate for the inflaton to the bulk modes. Our
scenario can be thought of as a hot radiating plate cooling down by
emitting its energy into its cold surrounding~\cite{Rouzbeh}.  To
estimate how efficient this process is, let us first comment on the
Kaluza-Klein (KK) decomposition of the bulk field. Consider the action
for a bulk scalar field, $\varphi$, of mass $\mu$, in the background
metric Eq.~(\ref{metric}),
\begin{equation}
  S_\chi = \int\! d^4x\, dz\, \sqrt{g^{(5)}}
  \left({1\over 2} g^{MN} \partial_M\varphi \partial_N\varphi -
  {1\over2} \mu^2\varphi^2 \right)~.
 \end{equation}
The equation of motion for the bulk scalar field then reads as 
 \be 
 \left[ - \partial_z^2 + 4\kappa~ {\rm sgn}(z) \partial_z + \mu^2 +
 \omega^{-2}(z)\partial^\mu\partial_\mu \right] \varphi(x,z) =0~.
 \ee
Since the fifth dimension is not compact, the KK spectrum is 
continuous, and the scalar field can be expanded as
 \be 
 \varphi(x,z) = \int {dm\over \kappa} \varphi_m(x) \chi(z;m)\,,
 \label{kkexp}
 \ee 
where the KK wave function, $\chi(z;m)$, satisfies the reduced field
equation; 
\begin{equation}
[-\partial_z^2 + 4\kappa\, {\rm sgn}(z) \partial_z + \mu^2
- \omega^{-2}(z)m^2 ] \chi(z;m) =0\,.
\end{equation}
The normalization condition is given by $\int dz\, \omega^2(z) \chi(z;m)
\chi(z;m') = \kappa \delta(m-m')$, while the boundary condition on the
brane is, $\partial_z\chi(z=0;m) =0$.

In the above equations, $m^2 = p^\mu p_\mu$ is the four dimensional mass.
The general solution to the above equation is given in terms of the
Bessel functions of index 
$\nu = \sqrt{4 + \mu^2/\kappa^2}$~\cite{Rubakov2,wise};
 \be 
 \chi(z;m) = {1\over N(m) \omega^2(z)}\, 
 \left[ J_\nu \left( {m\over \kappa\omega(z)}\right) 
 + A(m) Y_\nu \left( {m\over \kappa\omega(z)}\right) \right]\,,
 \label{kkchi}
 \ee
where $N$ is the normalization factor, and the coefficient $A(m)$ is
given by
 \be A(m) = -{2\, J_\nu(m/\kappa) + 
 (m/\kappa) J'_\nu(m/\kappa) \over 2\,
 Y_\nu(m/\kappa) + (m/\kappa) Y'_\nu(m/\kappa)}\,.  
 \ee
In the limiting case, where $\mu, m \ll \kappa$, one can take $\nu=2$
in the above expressions to approximate the KK wave function evaluated
at the brane position by, 
\be \chi(0;m)\approx \sqrt{\frac{m}{2}}~\,.
 \label{chi0}
\ee
Next we use these results to evaluate the decay rate of the inflaton
into the bulk modes. From Eq.~(\ref{bbc}), we can read the effective
coupling of the inflaton,
\be 
\xi\,[\chi(0;m)\chi(0;m')]~ \phi(x)\, \varphi_m(x)\, \varphi_{m'}(x)\,.  
\ee 
Note that this effective coupling constant is essentially proportional
to the geometric mean of the product of the KK masses, $\sqrt{mm'}$.
In our case the inflaton will preferably decay into those KK modes
with the largest possible fifth momentum allowed by the kinematics.
These excited modes will move away from the brane towards infinity.

The total decay rate is estimated in the low energy limit, where
$\mu,m_\phi\ll \kappa$, as
 \be  
   \Gamma_\phi = \int_{0}^{m_\phi}\int_{0}^{\sqrt{m_\phi^2 - m^2}}\,
            {dm\over \kappa}{dm'\over \kappa}\, \xi^2\,
        {[\chi(0,m) \chi(0,m')]^2 \over m_\phi}\,
     \approx
        {\xi^2\over 32} \left({m_\phi\over \kappa}\right)^2\, m_\phi \,.
 \label{gamma}
 \ee 
The above expression clearly shows that for a sufficiently heavy
inflaton, the decay rate to the bulk KK modes is quite efficient.

Let us now briefly discuss the fate of the energy density which is
stored in the bulk modes. From the four dimensional point of view one
can track the inflaton energy density from the Friedmann equation,
Eq.~(\ref{friedman}). The last contribution in the right hand side
actually accounts for a dark radiation, which depends on the redshift
factor; $(a_{h}/a)^4$. In fact this ratio becomes exceedingly small
during the exponential expansion of the Universe compared to the other
terms in the Friedmann equation. We can estimate this ratio by
assuming that at the beginning of inflation the energy density stored
in the vacuum energy density of the inflaton. For the time being we
also assume that there is no other source for energy density, in that
case from the brane observer point of view, the largest scale could be
the Compton wavelength corresponding to the inflaton mass scale, which
is at most: $a_{h}\sim [V(\phi)]^{-1/4}$. For typical
values, $V(\phi)^{-1/4}\sim 10^{-16}~({\rm GeV})^{-1}\sim
10^{-32}$~m., we find $a_{h}$ is negligible compared to the present
size of the Hubble horizon. The ratio, $a_{h}/a\ll 1$, becomes 
small by the time when inflation comes to an end.

In fact, in our case, we can roughly estimate the dark radiation
\footnote{This should not be confused by the primordial dark energy
contribution we referred to the unknown sector of the inflaton energy
density.}  contribution to the present day. We estimate the rate of
loss of energy
\begin{equation}
\frac{\Delta \dot\rho}{\rho}={\cal O}(1)\Gamma_{\phi}\,.
\end{equation}
The order one coefficient takes into account of the relativistic
degrees of freedom of the inflaton decay products. At sufficient late
times, $1/H\gg 1/\kappa $, or $\rho \ll {\rm
max}~\{V(\phi),~\sigma\}$, the total contribution to the dark radiation
will be roughly given by the loss in the energy density from the
brane, e.g., $\Delta\dot\rho+\Delta\dot\rho_{d}=0$,~\cite{March},
\begin{equation}
\Omega_{d}=\frac{\rho_{d}}{\rho_{d}+\rho}\approx 
\int_{\tau_{1}}^{\infty}~d\tau\left(-\frac{\Delta \dot\rho}
{\rho}\right)\,.
\end{equation}
We can estimate the integral and the final answer is given by
\begin{equation} 
\Omega_{d}\approx \frac{\xi^2\pi}{64}\left(
\frac{m_{\phi}}{\kappa}\right)^3\,.
\end{equation}
For $\kappa\gg m_{\phi}$, the dark radiation contribution is negligible.

There are other notable effects of the inflaton quanta escaping into he
bulk. Due to causality they will leave a wake in the form of
gravitational fluctuations. In fact, a particle moving with a
trajectory perpendicular to the brane follows a world line $z(t) =
\ln(1+ \kappa^2 t^2)/2\kappa$. As the particle leaves the brane, it
produces a gravitational wave, which dies away as, $\sim r/t$, for
distances $r$ within the light cone, see for instance
Ref.~\cite{Rubakov2} and the last Ref. in~\cite{Muck}.  These effects
will not be imprinted on the cosmic background, because these are
sub horizon effects which are diluted away in the course of expansion.
For our purposes we will
now continue working with the standard cosmology with $H\propto
\rho^{1/2}/M_{p}$.


\section{Neutrino masses}

Neutrino masses can well have its origin in four dimensions. Let us
assume a simple superpotential for the right handed Majorana
neutrinos,
\begin{equation}
\label{superpot}
W=\frac{1}{2}M_{N} {\bf N}~{\bf N}+h{\bf N}~{\bf L}~{\bf H_{u}}\,,
\end{equation}
where ${\bf N},~{\bf L},~{\bf H_{u}}$ stand for the neutrino, the
lepton and the Higgs doublet, $M_{N},~h$ are $3\times 3$ mass and the
Yukawa matrices respectively.  We have assigned an odd
$R$-parity for the right handed (s)neutrinos, which prohibits $({\bf
N})^3$ term in the superpotential. The only possible interactions are
through the Yukawa matrix, $h_{ij},~i,~j=1,2,3$, with the SM lepton
and the Higgs fields. The origin of the mass term $M_{N}$ could be due
to SO(10) broken down to SU(5) through the condensation of the SU(5)
singlet components in $(126,\overline{126})$ representation, for
instance. However the crucial point is that below the SO(10) breaking
scale, the D-term decouples from the potential of the sneutrino,
$\widetilde N$, which restricts the vev as, $\widetilde N\leq
M_{GUT}$.  Among other possibilities we can also think of breaking a
simple group, for e.g., $U(1)_{B-L}$, which can gives rise to the
masses of the neutrinos.

Irrespective of the origin of the right handed neutrino mass term, we
can write down a potential barring any non-renormalizable
interactions, and soft SUSY breaking $A$-terms,
\begin{equation}
V\sim M^2_{N}{\widetilde N}^2\,. 
\end{equation}
Non-renormalizable terms including higher powers of $N$ fields can
always be removed by the  addition of $R$-symmetry, which we assume,
whereas the mass scale of the soft supersymmetry breaking terms is
typically small (about few TeV) compared to the expected mass scale of
the right handed neutrinos, required to be around $10^{10}$~GeV by the
see-saw in our case.  For simplicity we always assume a diagonal basis
for the right handed neutrino mass matrix. The left handed neutrinos
obtain masses via a see-saw mechanism~\cite{Seesaw}
\begin{equation}
m_{\nu}\sim m_{D}^T M_{N}^{-1} m_D\,,
\end{equation}
where $m_{D}$ is the Dirac mass matrix.

For simplicity we may also assume that the lightest sneutrino mass to
be smaller than the Hubble expansion rate during inflation.  We also
assume that its largest Yukawa coupling is that of the $\tau$ doublet,
$h_{1,3}\sim 10^{-4}-10^{-5}$. Note that by taking $M_N\sim 10^{10}
GeV$, with $M_{N_{2,3}}\gg M_{N_{1}}$, we can obtain a small
electron-tau neutrino mixing in the left handed sector, as it seems to
be required by the small $\theta_{13}$ angle in the neutrino
experiments, provided there is a natural hierarchy in the neutrino
masses.

A possible texture that may give the desired bi-large mixing in the
left handed neutrino sector can be given by~\cite{mohapatra,goh}
\[ 
m_\nu\sim m_0 
           \left(\ba{ccc} \lambda & \lambda & \lambda\\ 
               \lambda & 1+\lambda & 1 \\ 
	       \lambda & 1 & 1 \ea \right)
\]
where $m_0 \approx \sqrt{\delta m^2_{atm}}$ and $\lambda$ is a small
parameter. This texture gives hierarchical neutrino masses, with mass
eigenvalues $m_3\approx m_0$ consistent with atmospheric; and $m_2
\approx \lambda m_0$, for solar oscillations, with a
smaller $m_1$. We can also obtain $\tan\theta_{atm}\approx 1$ and
large mixing angle required for explaining the solar neutrino
oscillations.  Although our intention is not to present a model for
the neutrino masses in this paper, but it is nice to see that the
physics involved in our discussion seems to provide a consistent
picture with the minimal requirement on some models for the neutrino
masses and mixings.


\section{Dynamics and density perturbations from $\widetilde N_1$}

During vacuum dominated inflation there is always quantum fluctuations
inspite of the fact that inflation wipes out all the
inhomogeneities. A massless bosonic field can have Brownian jumps of
order $\sim H/2\pi$~\cite{Linde}. However during one e-folding of
inflation the field becomes homogeneous in a Hubble patch and
therefore acts as a cosmic condensate. These jumps facilitate the
lighter fields to obtain non-vanishing vacuum expectation values
during inflation. The maximum amplitude of the light field can be
determined by; $V(\widetilde N)\leq V(\phi)_{inf}$, such that the
inflaton energy density remains dominating.  We denote the lightest
right handed sneutrino by, $\widetilde N_1$, assuming that it is
lighter than the Hubble expansion rate during inflation.

We are interested in tracking the classical and quantum evolution of
$\widetilde N_{1}$. In fact $\widetilde N_{1}$ acts as a curvaton
field~\cite{Sloth,Lyth,Moroi}. During inflation the energy density of
the curvaton field is sub-dominant.  Consequently quantum fluctuations
of this field lead to the isocurvature perturbations. Current CMB
measurements telling us that the density perturbations are mainly of
the adiabatic type~\cite{WMAP}.  Hence for the curvaton scenario to
work isocurvature perturbations have to be converted into the
adiabatic ones. Such a conversion takes place when the contribution of
the curvaton energy density, $\rho_{{\widetilde N}_1}$, to the total
energy density in the universe grows, i.e., with an increase of
\be
r = \frac{3 \rho_{{\widetilde N}_1}}{4 \rho_\gamma + 3 
\rho_{{\widetilde N}_1}}\label{r}\,.
\ee
Here $\rho_\gamma$ is the energy in the radiation bath from the
inflaton decay. Non-Gaussianity of the produced perturbation requires
the curvaton to contribute more than 1\% to the energy density of the
Universe at the time of decay, that is $r_{\rm dec} > 0.01$, required
by the present observation~\cite{WMAP}.  In our case the inflaton
energy density does not reheat the Universe, therefore, the above
ratio, $r=1$, is naturally satisfied, suggesting that there is no
possible signature of primordial non-Gaussianity in CMB
fluctuations. This should be considered as one of our predictions.

The long wavelength fluctuations in the sneutrino field leaves the
Hubble horizon, these perturbations can be defined on a finite energy
density hypersurface foliated in a coordinate system, such that the
metric perturbation is $\zeta$ and the metric (for a detailed
discussion on cosmological density perturbations, see~\cite{Brandenberger})
is given by,
\begin{equation}
ds^2= a^2(t)\left(1+2\zeta \right)\delta_{ij}dx^{i}dx^{j}\,,
\end{equation}
where $a$ is the scale factor. The time evolution of the curvature
perturbation, $\zeta$, on scales larger than the size of the horizon is
given by~\cite{Bardeen,Kodama,Gordon}
\begin{equation}
\label{curv0}
\dot\zeta=-\frac{H}{\rho+P}\delta P_{{\rm nad}}\,,
\end{equation}
where $P_{{\rm nad}}\equiv \delta P-c_{s}^2\delta\rho$ is the
non-adiabatic pressure perturbation. The adiabatic sound speed is
$c_{s}^2=\dot P/\dot\rho$, where $P$ and $\rho$ are the total pressure
and energy density, respectively. For a single field $\delta P_{{\rm
nad}}=0$, therefore on large scales the curvature perturbation is
pure adiabatic in nature with $\zeta={\rm constant}$. This is not true
in presence of many fields, because relative pressure perturbations
between fields can give non-zero contribution to $\delta P_{\rm nad}$.

The unperturbed and the perturbed equations of motion for the
sneutrino are given by,
\begin{eqnarray}
\ddot{\widetilde N}_1+3H\dot{\widetilde N}_1+M_{N,1}^2N_{1}=0\,,\\
\delta\ddot{\widetilde N}_{1k}+3H\dot\delta{\widetilde N}_{1k}+\frac{k^2}{a^2}
\delta{\widetilde N}_{1k}+M_{N,1}^2\delta{\widetilde N}_{1k}=0\,.
\end{eqnarray}
In the limit when $M^2_{N,1}\ll H^2$, the perturbations in $\widetilde
N_1$ are Gaussian, and the spectrum is given by
\begin{equation}
{\cal P}_{N_1}^{1/2}=\frac{H_{\ast}}{2\pi}\,,
\end{equation}
where $\ast$ denotes the epoch of horizon exit $k=a_{\ast}H_{\ast}$.

After the energy density stored in the sneutrino condensate overtakes
the redshifted inflaton energy density, the sneutrino condensate 
starts oscillating. During this epoch the mean squared fluctuations in
the sneutrino condensate follows:
\begin{equation}
\langle(\delta N_1)^2\rangle=\int_{k_{min}}^{k_{max}}{\cal P}_{N_1}(k)
\frac{dk}{k}\sim \frac{H_{\ast}^2}{4\pi^2}\ln
\left(\frac{k_{max}}{k_{min}}\right)\sim \frac{H_{\ast}^2}{4\pi^2}\,.
\end{equation}
where $k_{max}\sim (\tilde a\tilde H)$ and tilde denotes the
sub horizon redshifted modes and $k_{min}=a_{0}H_{0}$, where $0$
corresponds to the present values. The field perturbations is then 
given by
\begin{equation}
\delta=2\frac{\delta N_1}{N_1}=\frac{H_{\ast}}{\pi N_{1\ast}}\,,
\end{equation}
when $N_{1,\ast}\sim 10^{5}\times H_{\ast}$, in order to match the
amplitude for the CMB fluctuations. On the other hand we require the
lightest sneutrino mass should be smaller than the Hubble expansion
rate during inflation, 
\begin{equation}
    V''({\widetilde N}_*) = M^2_{N_1}\sim \alpha^2 H_*^2\,,
\end{equation}
where $\alpha< 1$, while the scale of inflation is given by,
$V_{I}^{1/4}\sim(H_*M_p)^{1/2}$.

The amplitude of the fluctuation of the sneutrino field created during
inflation will be imprinted on radiation when the condensate
decays. The total curvature perturbations is then given by~\cite{Lyth}
\begin{equation}
\zeta=\zeta_{N_1}=-H\frac{\delta\rho_{N_1}}{\rho_{N_1}}=\frac{1}{3}\delta\,,
\end{equation}
and ${\cal P}^{1/2}_{\zeta}\approx H_{\ast}/N_{1\ast}$.  The
spectral index of microwave temperature perturbations can then be
evaluated as \cite{LyWa}
\begin{equation}
\label{nspectr}
    n_s-1 = \frac{d\ln {\cal P}_{N_1}}{d\ln k}=2 \frac{\dot{H_*}}{H_*^2}
    + \frac{2}{3}\frac{M_{N_1}^2}{H_*^2}\sim \frac{2}{3}\alpha^2\,.
\end{equation}
For $\alpha \sim 10^{-1}$, we find the spectral index is fairly close
to one, which is consistent with the recent WMAP observation of
$n_s=0.99 \pm 0.04$~\cite{Spergel}. The only assumption is that the
Hubble expansion rate during inflation is almost constant, therefore
the first term in the spectral index is negligible while the main
contribution comes from the curvature of the sneutrino field.

Since the sneutrino field is dominating the energy density, while
decaying into sleptons and Higgsinos, there is no residual
isocurvature perturbations left over, all of them are converted into
the adiabatic ones. Later on we will comment on the
baryon-isocurvature perturbations.


\section{Reheat Temperature}

In our case it is the coherent oscillations of the sneutrino field
which leads to reheating. For the details of sneutrino reheating we
refer the readers to Ref.~\cite{Marieke}. In our case the perturbative
decay rate of the lightest sneutrino condensate can be estimated by,
$\Gamma\sim |h_{13}|^2M_{N_1}/4\pi$, and the reheat temperature,
$T_{rh}\sim 0.1\sqrt{\Gamma M_{p}}$,
\begin{equation}
T_{rh}\sim 10^{9}\left(\frac{h_{13}}{10^{-4}}\right)\left(\frac{M_{N_1}}{
10^{10}}\right)^{1/2}~{\rm GeV}\,.
\end{equation}
In order to avoid the gravitino problem, the reheat temperature of the
universe must be $10^{9}$~GeV \cite{Ellis}. In order to satisfy this
bound, the lightest right handed sneutrino (neutrino) mass must be
around $10^{10}$~GeV. This constraints the scale of inflation to be
$H_{\ast}\sim 10^{10}/\alpha$~GeV. If $\alpha\sim 10^{-1}$, then
$H_{\ast}\sim 10^{11}$~GeV and the initial sneutrino vev, $\langle
{\widetilde N}_1\rangle \sim 10^{16}$~GeV, which is close to the grand
unification scale.

The above numbers are encouraging, because we can nail down the mass
of the lightest right handed neutrino and the initial vev for the
sneutrino field.  The latter may be indicative that the origin of the
masses to the right handed neutrinos is linked to the grand unified
theories (GUT), as for instance $SO(10)$ GUT models.  It would be
interesting to further explore this possibility.  Further note that
the sneutrino vev is smaller than the $SO(10)$ breaking scale,
therefore justifying our approximation to ignore the D-term
contributions in the sneutrino superpotential, see
Eq.~(\ref{superpot}). A further comment on the neutrino masses is that
the right handed neutrino mass scale might be different than the GUT
scale, provided it originates via non renormalizable operators, as it
happens in many $SO(10)$ constructions.

\section{Baryogenesis}

Let us discuss baryogenesis. The sneutrino decay also induces lepton
asymmetry due to the CP violation. Net CP asymmetry in our scenario
can be calculated by computing the interference between the tree-level
and the one loop diagrams of $\widetilde N$ going into leptons,
$\ell_{j}{\widetilde H_{u}}$ (see figure 1), and into antisleptons,
$\tilde{\ell}_{j}^{\ast}H_{u}^{\ast}$ (see figure 2 ), where $j=1,2,3$
flavors~\cite{Fukugita,Murayama}, including the contribution from the
wave function mixing diagrams like those depicted in Figure
3~\cite{covi}.  Notice that typical 1-loop diagram that involve
quartic scalar couplings do appear from F-terms of the potential,
however they do not contribute to CP asymmetry, although they are
needed to cancel the quadratic anomalies.
 
\begin{figure}[ht]
\centerline{
\epsfxsize=250pt
\epsfbox{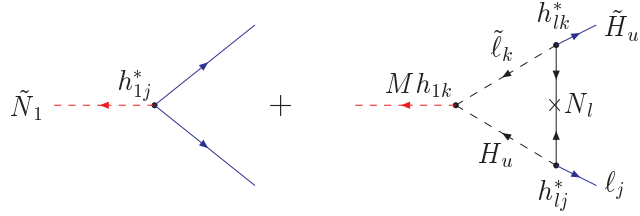}
}
\vskip1ex

\caption{Tree level and one loop diagrams for the $\tilde N_1$
decaying into lepton and Higgsino, $\ell_j \tilde{H}_u$. }
\end{figure}

\begin{figure}[ht]
\centerline{
\epsfxsize=250pt
\epsfbox{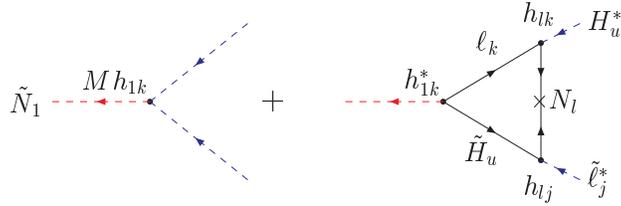}
}
\vskip1ex

\caption{Tree level and one loop diagrams for the decay of $\tilde N_1$  into
antislepton and Higgs, $\tilde{\ell}_j^\ast H_u^\ast$. }
\end{figure}

\begin{figure}[ht]
\centerline{
\epsfxsize=250pt
\epsfbox{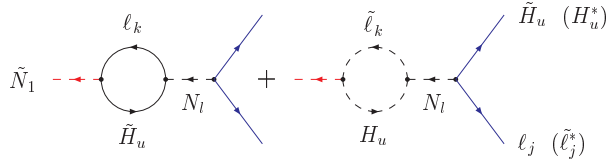}
}
\vskip1ex

\caption{One loop--type diagrams that contribute to wave function
mixing for $\tilde N_1$ decaying into $\ell_j \tilde{H}_u$.  Similar
diagrams exist for the decay into $\tilde{\ell}_j^\ast H_u^\ast$, with
the corresponding exchange of external lines as indicated.}
\end{figure}

The CP asymmetry in each decay channel for $X$ (where $X=
\ell_{j}{\widetilde H_{u}};~\tilde{\ell}_{j}^{\ast}H_{u}^{\ast}$) is
defined as 
\be 
\epsilon_X \equiv \frac{\Gamma_{\widetilde{N}^\ast X} -
\Gamma_{\widetilde{N}\bar X}} {\Gamma_{\widetilde{N}^\ast X} +
\Gamma_{\widetilde{N}\bar X}}~.  
\ee 
A direct computation of  the total CP asymmetry gives~\cite{covi} 
\be 
\epsilon= -\frac{1}{4\pi}\sum_{k\neq 1} \left[
\left(\frac{M_{N_k}}{M_{N_1}}\right)
\ln\left(\frac{M_{N_1}^2}{M_{N_k}} + 1\right)+
2\frac{M_{N_1}M_{N_k}}{M_{N_k}^2-M_{N_1}^2} \right]{\cal I}_{k1}\,,
\ee
arising from the vertex and the wave function CP violating diagrams,
respectively, where
\be 
{\cal I}_{k1} =
\frac{Im\left(h_{1j}h_{1k}h_{lj}^\ast
h_{lk}^\ast\right)}{h_{1j}h_{1j}^\ast}\,. 
\ee 
In the limit of hierarchical right handed masses, $M_{N_1}\ll
M_{N_{2,3}}$, which we are considering, we obtain a simplified
expression for the CP asymmetry~\cite{covi} 
\be 
\epsilon=
-\frac{3}{4\pi}\sum_{k\neq 1}
\left(\frac{M_{N_1}}{M_{N_k}}\right){\cal I}_{k1}\,. 
\ee 
The suppression factor, $M_{N_1}/M_{N_k}$, implies that small CP
asymmetry is likely to be produced.

Now we can predict the left handed neutrino mass scale with the help
of seesaw formula, within our hypothesis of $h_{33}$ dominance,
$m_{\nu}=|h_{33}|^2{\sin^2\beta}/{(2\sqrt{2}G_{F}M_{N_3})}$, where
$G_{F}$ is the Fermi constant and $\tan\beta=\langle
H_{u}\rangle/\langle H_{d}\rangle$. If we take $|h_{33}|\sim
10^{-1}$ with a typical $\tan\beta\sim 10$ and $M_{N_3}\sim
7\times 10^{13}$~GeV, we obtain $m_3\sim 0.04$~eV, which is already
around the expected mass scale for the neutrinos in the case of
hierarchical neutrino masses.

The lepton asymmetry can be estimated from the Boltzmann equation.
\begin{equation}
\dot n_{L}+3Hn_{L}=\epsilon\Gamma\frac{\rho_{N_1}}{M}-\frac{8}{23}
\frac{G_{F}^2m_{\nu}^2}{\sin^4\beta}T^3n_{L}\,.
\end{equation}
The integration becomes simple especially since we can neglect the
rescattering rate of produced leptons which is proportional to the squared
mass of the neutrino.   
The lepton asymmetry is finally converted by the SM sphalerons into
the baryons with a net baryon number to the entropy, which is given
by~\cite{Murayama,covi}
\begin{equation}
\frac{n_{B}}{s} \sim \epsilon \frac{6}{23}\frac{T_{rh}}{M_{N_1}}\,.
\end{equation}
The factor $T_{rh}/M_{N_1}$ arises due to the entropy generation from
the decay of the condensate.  We note that with $\epsilon\sim
10^{-8}$, we can easily generate baryon asymmetry of order one part in
$10^{10}$, for $T_{rh}\sim 10^{9}$~GeV and $M_{N_1}\sim
10^{10}$~GeV. The actual prediction is an interplay between $T_{rh}$,
$M_{N_1}$ and $\epsilon$.  Our analysis seems to favor rather small
value for~$\epsilon$.

The fluctuations in the reheat temperature also lead to the
fluctuations in the baryon asymmetry, $\delta(n_{B}/s)\propto \delta
T_{rh}$. However in our case these fluctuations are adiabatic in
nature.  There is no residual baryon-isocurvature fluctuations
generated during the sneutrino decay, because the condensate dominates
the energy density while decaying, therefore converting all its
isocurvature fluctuations into the adiabatic ones. In our simple
set-up we predict, $S_{B}=0$. This should be taken as a robust claim
which can be ruled out very easily from future microwave background
anisotropy measurements.

\section{Conclusion}

As far as the present state of art goes, the origin of the inflaton,
its potential and its couplings to the SM are unknown. For various
reasons the inflaton is often treated as a gauge singlet, on the other
we also often `assume' that the inflaton energy density is dumped into
the SM or MSSM degrees of freedom.  Here we do not make this
assumption; rather we examine what would happen if the inflaton energy
density were not to reheat the MSSM (SM) degrees of freedom?

In this regard we separate the two sectors, one of which is
responsible for inflation, and the other is responsible for generating
density perturbations, reheating the Universe with the MSSM degrees of
freedom, and creating the observed baryon asymmetry. Our main
observation is that there is no logical necessity for the inflaton
energy density to be responsible for generating the hot big bang
cosmology. The entire thermal history of the Universe can be created
from the decay of the sneutrino condensate. It is quite interesting to
see that we obtain the right amplitude for the density perturbations,
the neutrino mass scale, the reheat temperature below the gravitino
overproduction bound, and the observed baryon asymmetry without much
fine tuning. An important input in the whole setup is the sneutrino
vev, which appears to be close to the grand unification scale.


\section*{Acknowledgments}

We would like to thank Kari Enqvist, James Cline and Horace Stoica 
for fruitful discussions.  A.M. is a CITA-National fellow. APL's work
is supported in part by CONACyT, M\'exico, under grant J44596-F.


\end{document}